\documentclass[preprint,review,12pt]{elsarticle}

\usepackage{amssymb}

\journal{Physica B: Condensed Matter}

\begin{document}

\begin{frontmatter}



\title{An approach for studying the influence of uniaxial strain (pressure) on the
temperature of the Bose-Einstein condensation of intersite bipolarons:
possible implementation for RBa$_2$Cu$_3$O$_{7-\delta}$ cuprates}


\author[els1,els2]{B.Ya. Yavidov\corref{cor1}\fnref{fn1}}
\ead{yavidov@inp.uz}

\author[els1]{Sh. S. Djumanov}

\author[els1]{E. Karimboev}

\cortext[cor1]{Corresponding author} \fntext[fn1]{Institute of
Nuclear Physics, 100214 Ulughbek, Tashkent, Uzbekistan\\ Tel. +998
71 289 25 57/ Fax: +998 71 150 30 80}

\address[els1]{Institute of Nuclear Physics, 100214 Ulughbek, Tashkent, Uzbekistan}
\address[els2]{Nukus State Pedagogical Institute named after A'jiniyaz, 230105 Nukus, Karakalpakstan, Uzbekistan}

\begin{abstract}

A universal approach is proposed to study the influence of strain
(pressure) on the temperature of Bose-Einstein condensation of
intersite bipolarons within the extended Holstein model. It is shown
that uniaxial strain (pressure) derivatives of the temperature of
such a Bose-Einstein condensation strongly depend on the arrangement
of ions in the lattice. In particular, they may be positive or
negative. A connection between the theoretically obtained results,
along with the experimental data, on the influence of uniaxial
pressure (strain) on $T_c$ of RBa$_2$Cu$_3$O$_{7-\delta}$ family
cuprates is discussed.

\end{abstract}


\end{frontmatter}



\section{Introduction}

The extended Holstein model (EHM) was introduced in Ref.
\cite{alekor}, primarily for the explanation of smallness of charge
carrier's mass in the cuprates. In the strong coupling regime (or
when the electron-phonon interaction is strong), the small value of
EHM polaron's mass compared to that of the usual Holstein model
\cite{hol}, is primarily ensured by the long-range nature of the
electron-phonon interaction (interaction via Fr\"{o}hlich type
density-displacement force). In the last decade, different
properties of the EHM were studied by several authors (see review
article \cite{devr-asa}). In Ref.~\cite{kor-ctqmc}, Kornilovitch
studied the ground state energy, effective mass and polaron spectrum
with the help of the continuous-time Quantum Monte Carlo algorithm.
An anisotropy of the polaron's mass due to electron-phonon
interaction, ground-state dispersion, and density of states of the
EHM polaron were studied in Refs.\cite{kor-giant,kor-ground}.
 Fehske, Loos and Wellein \cite{flw} investigated the electron-lattice
 correlations, single-particle spectral function and optical conductivity of a
polaron within the EHM in the strong and weak coupling regimes by means of
  the Lanczos diagonalization method. Other properties of the EHM, such as the
ground state spectral weight, the average kinetic energy and the mean
  number of phonons were studied in Refs.\cite{pcf-jpcm,cfmp-prb,hohen}
   by means of the variational and Quantum Monte Carlo simulation approaches.
  The work of Ref.~\cite{asa-ya} extended the EHM to the adiabatic limit.
The influence of the different types of polarized vibrations, and the
arrangement of the ions on the mass of polaron, have been studied in
Refs.\cite{trg,yav-jetp,yav-physb}.
The EHM with screened electron-phonon interactions was discussed in
  Refs. \cite{kor-giant,spencer,hague-etal,hague-kor,yav-pla,yav-epjb}.
The influence of both local and nonlocal electron-phonon interactions
on the full polaronic effect within the EHM was studied in Ref. \cite{stojan},
finding that these interactions can compensate for each other,
resulting in the suppression of polaronic effects. The EHM was used
to study a polaron formation in semiconducting polymers,
 finding the polaron's energy, its size, and lattice
deformation as a function of the conjugation length \cite{meisel}.

Quite recently, one of us (B.Ya.Ya.) extended an application of the
EHM to La-based cuprate films under pressure (strained films)
\cite{yav-physc-strain}. Namely, in that work, a unified approach
for studying the influence of pressure (stress or strain) on the
temperature of Bose-Einstein condensation of intersite bipolarons
was proposed. Uniaxial strain derivatives of Bose-Einstein
condensation temperature ($T_{BEC}$) of intersite bipolarons were
calculated. Having accepted the bipolaronic scenario as a ground for
high-T$_c$ superconductivity of cuprates, the experimental results
on the influence of lattice mismatch to the critical temperature
($T_c$)
 of La-based high-T$_c$
films were explained. In particular, the results of two experiments
\cite{sato-naito,locquet} were explained within the framework of the
EHM and the bipolaronic theory of superconductivity. The main
features of the model proposed in Ref.~\cite{yav-physc-strain} are:
(i) compressive pressure (strain) in the $ab$--plane of cuprates
enhances $T_{BEC}$ and (ii) compressive pressure (strain) along
$c$-axis of cuprates reduces $T_{BEC}$. Such variations of the $T_c$
of cuprates with respect to applied pressure (strain) are often
observed \cite{schilling-handbook}. The proposed model allows one to
interrelate strains in each axis with each other and to study their
cumulative effect on $T_{BEC}$.

Here, we continue to study the influence of strain induced by
external pressure or lattice mismatch on $T_{BEC}$ for different
lattices. Special attention will be given to the possibility of
qualitative explanation of such phenomena as sign difference of the
strain (pressure) derivatives of $T_c$ of
RBa$_2$Cu$_3$O$_{7-\delta}$ cuprates (R stands for Yb, Y, Dy or Gd)
along the $a-$ and $b-$ axes. RBa$_2$Cu$_3$O$_{7-\delta}$ compounds
stand apart from other cuprates (with the CuO$_2$ planes only)
because of the presence of Cu-O chains along the $b$-axis in the
crystal structure. Strong anisotropy of crystal structure gives rise
to the anisotropy of electronic, thermodynamic, transport and other
properties of the RBa$_2$Cu$_3$O$_{7-\delta}$ family of materials.
In particular, uniaxial strain derivatives of the critical
temperature, $T_c$, along crystallographic axes $a$ and $b$  have
opposite sign: $\partial T_c/\partial\varepsilon_a<0$ and $\partial
T_c/\partial\varepsilon_b>0$. The value of the uniaxial strain
derivative of the critical temperature of cuprates along $c-$ axis,
$\partial T_c/\partial\varepsilon_c$, lies in a wide range, but all
values are negative. There are several works that theoretically
explain uniaxial the strain (pressure) derivatives of the critical
temperature of RBa$_2$Cu$_3$O$_{7-\delta}$ cuprates
\cite{millis-rabe,q.p.li,klein-simanovsky,pickett,jansen-block,chen-habermeier}.
Though the theoretical models (see above works) proposed so far were
able to somehow explain the uniaxial pressure (strain) derivatives
of $T_c$ of RBa$_2$Cu$_3$O$_{7-\delta}$ compounds and reproduce some
aspects of the well-known experiments \cite{meingast-prl-91,
welp-prl-69}, they miss the important issue relevant to all
cuprates. The issue is the presence of strong electron-phonon
interaction in the cuprates
\cite{gunnar-rosch-jpcm-2008,mish-phys.usp-2009} and the polaronic
nature of charge carriers
\cite{zhao-Pol.Adv.Mat,BussHol-Keller-Pol.Adv.Mat}. An another point
is that those models are applicable to only one cuprate compound,
for example to the RBa$_2$Cu$_3$O$_{7-\delta}$ family, and do not
discuss the same problems in La- and Bi-based cuprates. In this
sense those models lack universality. Therefore, at present,
development of a model that explains the uniaxial strain (pressure)
derivatives of $T_c$ from the universal point of view and takes into
account strong electron-phonon interaction in the cuprates is one of
the most important research tasks in the way of our understanding
the microscopic origin of high-$T_c$ phenomena.


Our model, proposed in  \cite{yav-physc-strain}, does not suffer
from the above-mentioned imperfections. Below we show that the EHM
and Bose-Einstein condensation scenario of intersite bipolarons, if
one accepts the latter to be responsible for high-T$_c$
superconductivity of cuprates, are able to qualitatively explain the
sign difference of $\partial T_c/\partial\varepsilon_a<0$ and
$\partial T_c/\partial\varepsilon_b>0$.

\section{The Model Hamiltonian and lattices}

We write the Hamiltonian of the system of electrons and phonons
\cite{alekor,asa-kor-jpcm} as
\begin{equation}\label{Ham-full}
H=H_{e}+H_{ph}+H_{V}+H_{e-ph},
\end{equation}
where
\begin{equation}\label{H-e}
H_{e}=\sum_{{\bf n}\neq {\bf n'}}T({\bf n}-{\bf n'})c^{\dag}_{\bf
n}c_{\bf n'}
\end{equation}
describes the hopping of electrons between adjacent sites,
\begin{equation}\label{H-ph}
H_{ph}=\sum_{{\bf q},{\alpha}}\hbar\omega_{{\bf
q}\alpha}\left(d^{\dag}_{\bf q\alpha}d_{{\bf q}\alpha}+1/2\right),
\end{equation}
is the Hamiltonian of the phonon system,
\begin{equation}\label{H-Vc}
H_{V}=\sum_{{\bf n}\neq {\bf n'}}V_{C}({\bf n}-{\bf
n'})c^{\dag}_{\bf n}c_{\bf n}c^{\dag}_{\bf n'}c_{\bf n'},
\end{equation}
is the Hamiltonian of interacting particles at sites $\bf n$ and $\bf
n'$ via Coulomb forces, and
\begin{equation}\label{H-e-ph}
H_{e-ph}=\sum_{{\bf n}{\bf m}\alpha}f_{{\bf m}\alpha}({\bf
n})c^{\dag}_{\bf n}c_{\bf n}\xi_{{\bf m}\alpha}
\end{equation}
is the Hamiltonian of electron-phonon interaction. Here $T(\bf n-\bf
n')$ is the transfer integral of electron from site $\bf n$ to site
$\bf n'$, $c^{\dag}_{\bf n}(c_{\bf n})$ is the creation
(annihilation) operator of an electron at site $\bf n$,
$d^{\dagger}_{\bf q\alpha}(d_{\bf q\alpha})$ is the creation
(annihilation) operator of a phonon with $\alpha$ ($\alpha=x,y,z$)
polarization and wave vector $\bf q$, $\omega_{{\bf q}\alpha}$ is
the phonon's frequency, $V_{C}(\bf n-\bf n')$ is the Coulomb
potential energy of two electrons located at sites $\bf n$ and $\bf
n'$, $f_{\bf m\alpha}(\bf n)$ is the ''density-displacement'' type
coupling force of an electron at site ${\bf n}$ with the apical ion
at site ${\bf m}$ (Fig.1), and $\xi_{{\bf m}\alpha}$ is the normal
coordinate of ion's vibration on site ${\bf m}$ which is expressed
through phonon creation and destruction operators as
\begin{equation}\label{xi}
    \xi_{{\bf m}\alpha}=\sum_{\bf q}\left(\sqrt{\frac{\hbar}{2NM\omega_{{\bf
    q}\alpha}}}e^{i{\bf qm}}d^{\dagger}_{{\bf
    q}\alpha}+h.c.\right).
\end{equation}
Here $N$ is the number of sites and $M$ is the ion's mass. We work
with dispersionless phonons and take into account only the
$c$-polarized vibrations of ions, as charge carriers in the CuO$_2$
plane of the cuprates strongly interact with $c-$ polarized
vibrations of apical ions \cite{timusk}. The lattice of Figure 1A
was introduced by Alexandrov and Kornilovitch in Ref. \cite{alekor}
in order to mimic the interaction of a hole on the CuO$_2$ plane
with the vibrations of {\it apical} ions in the cuprates (Figure 1B
is considered in Ref. \cite{bt}). Such one dimensional lattices have
similarity with some ions arrangements in the real structure of
RBa$_2$Cu$_3$O$_{7-\delta}$ cuprates. Indeed in the both lattices
the upper chain (open circles) represents apical ions of oxygen at
position O(4). The lower chain of Fig.1A consists of copper Cu(1)
ions of Cu(1)-O(1) chain or copper Cu(2) ions of CuO$_2$ plane.
Fig.1B resembles arrangement of oxygen O(2) ions of CuO$_2$ plane
and O(4) oxygen ions.

The occurrence of an anisotropy of charge carrier's mass in
RBa$_2$Cu$_3$O$_{7-\delta}$ compounds we model here by the
orientational dependence of the "density-displacement" type
electron-phonon interaction on the relative positions of out of
plane ions with respect to copper-oxygen (CuO$_2$) plane. The latter
effect is caused by the anisotropy of the structure
RBa$_2$Cu$_3$O$_{7-\delta}$ of compounds. In this way we can model
an interaction of the hole belonging to copper-oxygen plane with the
apical oxygen ions at O(4) positions. Though such a primitive model
lattices far from the real structure of RBa$_2$Cu$_3$O$_{7-\delta}$
compounds, we will see below that they are able to capture the
essential physics of the influence of strain (pressure) to T$_c$ of
RBa$_2$Cu$_3$O$_{7-\delta}$ compounds.

\begin{figure}[tbp]
\begin{center}
\includegraphics[angle=-0,width=0.75\textwidth]{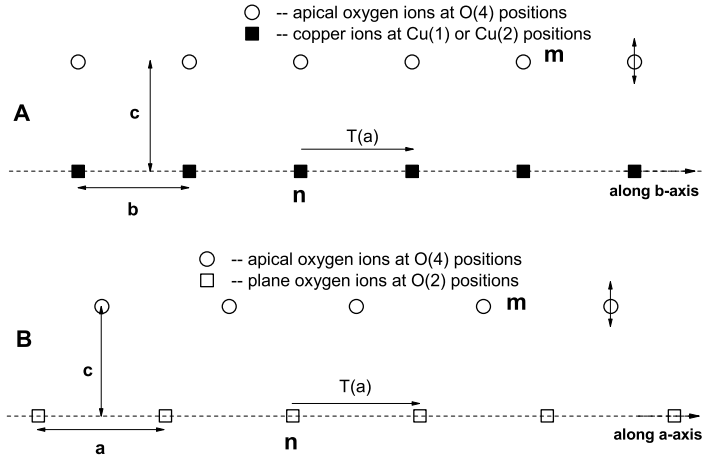} \vskip -0.5mm
\end{center}
\caption{An electron hops on a lower chain of ions (squares) and
interacts with the $c$--polarized vibrations of ions (open circles)
of an upper chain, via a density-displacement type of force $f_{\bf
m}({\bf n})$. The distances between the chains ($|{\bf c}|$) and
between the ions ($|{\bf a}|$ or $|{\bf b}|$) of the same chain are
assumed equal to 1.}
\end{figure}

It has been shown that within the model of Eq.~(\ref{Ham-full})
intersite bipolarons tunnel in the first order of polaron tunneling
and mass of the intersite bipolaron has the same order as polaron's
mass \cite{asa-kor-jpcm}. For the sake of simplicity we suppose here
that intersite bipolarons form an ideal gas of charge carriers and
mass of bipolaron is $m_{bp}=2m_p$ (this point does not lead to loss
of generality). Then the temperature of Bose-Einstein condensation
of the intersite bipolarons is defined as
\begin{equation}\label{t-bec}
 T_{BEC}=\frac{3.31\hbar^2n^{2/3}}{2k_Bm^{\ast}}e^{-g^2}.
\end{equation}
Here $k_B$ is Boltzmann constant and $n$ is density of intersite
bipolarons. For the ideal gas of charge carriers one can neglect
interparticle Coulomb interaction, i.e. the term $H_V$
(Eq.(\ref{H-Vc})). In this case one can estimate polaron's mass
within EHM. In the strong electron-phonon coupling limit and
nonadiabatic regime, use of the standard procedure such as
Lang-Firsov transformation \cite{lang-fir} eliminates
electron-phonon interaction term $H_{e-ph}$ (Eq.(\ref{H-e-ph})).
Subsequent perturbation expansion of the transformed Hamiltonian
with respect to the parameter $\lambda^{-1}=2T(a)/E_p$ ($E_p$-
polaron shift) and estimation of the polaron's renormalized mass
yields $m_p/m^{\ast}=\exp[g^2]$ \cite{alekor} (see also
\cite{asa-ya}), where
\begin{equation}\label{g2}
    g^2=\frac{1}{2M\hbar\omega^3}\sum_{\bf m}[f^2_{{\bf m}}({\bf n})-f_{{\bf
m}}({\bf n})f_{{\bf m}}({\bf n}+{\bf a})].
\end{equation}
and $m^{\ast}=\hbar^2/2T(a)a^2$ is the bare band mass. In order to
consider the stress of a lattice and its influence on the (bi)polaron
mass, and consequently on the temperature of Bose-Einstein
condensation of intersite bipolarons, the analytical expression
\begin{equation}\label{force-epsilon}
  f_{\bf m}({\bf n})=\frac{\kappa c(1-\varepsilon_c)}{[|({\bf n}-{\bf
m})(1-\varepsilon_{i})|^2+(c(1-\varepsilon_c))^2]^{3/2}}
\end{equation}
will be used for the density-displacement type force ($i=a$ or $b$).
Here $\kappa$ is some coefficient, and $\varepsilon_a$,
$\varepsilon_b$ and $\varepsilon_c$ are lattice strains along the
$a$-, $b$- and $c$-axes, respectively. The distance $|{\bf n}-{\bf
m}|$ is measured in units of the lattice constant $|{\bf a}|=1$ (or
$|{\bf b}|=1$). The lattice strains are defined as
$\varepsilon_a=(a_{unst}-a_{str})/a_{unst}$
($\varepsilon_b=(b_{unst}-b_{str})/b_{unst}$) and
$\varepsilon_c=(c_{unst}-c_{str})/c_{unst}$, where subscripts {\it
unstr} and {\it str} stand for unstrained and strained,
respectively. Eq.(\ref{force-epsilon}) is a generalization of the
force considered in Ref. \cite{alekor} (see Eq.(9) therein)  and
allows one to interrelate the temperature of Bose-Einstein
condensation of the intersite bipolarons with the lattice strains
through the mass of the intersite bipolaron.

\begin{figure}[tbp]
\begin{center}
\includegraphics[angle=-0,width=0.75\textwidth]{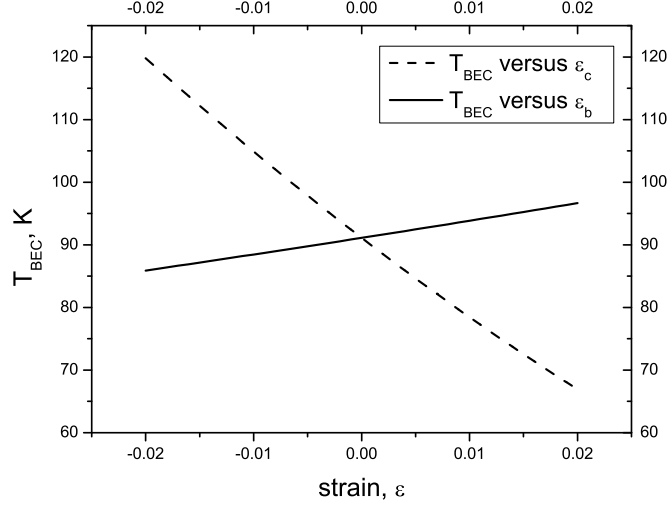} \vskip -0.5mm
\end{center}
\caption{The temperature of Bose-Einstein condensation of the
intersite bipolarons as a function of strains along the $b$-axis
$\varepsilon_b$ (solid line) and along the $c$-axis $\varepsilon_c$
(dashed line) for the lattice of Fig.1A. Here we put
$\kappa^2/(2M\hbar\omega^3)=5.885$ in order to coincide $T_{BEC}$ at
$\varepsilon_i=0$ ($i=b,c$) with the bulk value of $T_c\approx 91$ K
 of YBCO high-T$_c$ cuprates at optimal doping.}
\end{figure}

\begin{figure}[tbp]
\begin{center}
\includegraphics[angle=-0,width=0.75\textwidth]{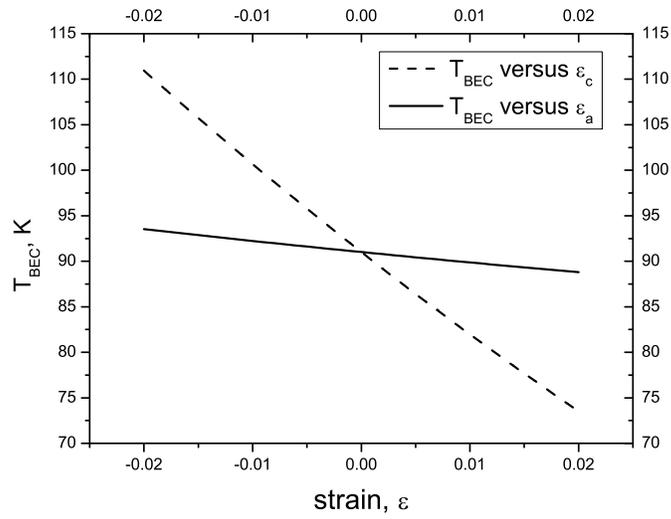} \vskip -0.5mm
\end{center}
\caption{The temperature of Bose-Einstein condensation of the
intersite bipolarons as a function of strains along $a$-axis
$\varepsilon_a$ (solid line) and along $c$-axis $\varepsilon_c$
(dashed line) for the lattice of Fig.1B. Here we put
$\kappa^2/(2M\hbar\omega^3)=9.265$ in order to coincide $T_{BEC}$ at
$\varepsilon_i=0$ ($i=a,c$) with the bulk value of $T_c\approx 91$ K
 of YBCO high-T$_c$ cuprates at optimal doping.}
\end{figure}

\section{Results and discussion}

The expression Eq.(\ref{t-bec}) expresses $T_{BEC}$ through two
basic parameters of a system: (i) the density of intersite
bipolarons $n$ and (ii) the exponent $g^2$ of the polaron mass
enhancement. Eq.(\ref{t-bec}) allows one to study the dependence of
$T_{BEC}$ on the model lattices (Fig.1) strains $\varepsilon_a$,
$\varepsilon_b$ or $\varepsilon_c$ at constant $n$. This dependence,
of course, originates from polaronic effects. We have calculated the
values of $T_{BEC}$ as a function of the strains along the $b$-axis
$\varepsilon_b$ and $c$-axis $\varepsilon_c$ for the model lattice
given in Fig.1A. The results are given in Fig.2. Here we put
$n=1\cdot 10^{21}$ sm$^{-3}$ and $\kappa^2/(2M\hbar\omega^3)=5.885$
in order to coincide $T_{BEC}$ in the absence of the strains with
the bulk value of $T_c\approx 91$ K of YBCO cuprates. The results
calculation of $T_{BEC}$ for the lattice in Fig.1B are shown in
Fig.3. As one can see from Fig.2, compressive strain along $b-$ axis
gives rise to increase the value of  $T_{BEC}$, while that along
$c-$ axis acts on the contrary. Compressive strain along both the
$a$- and $c$-axes in the model lattice of Fig.1B lowers the value of
$T_{BEC}$. The uniaxial strain derivatives of $T_{BEC}$ for the
model lattice given in Fig.1A at $\kappa^2/(2M\hbar\omega^3)=5.885$
are: $\partial T_{BEC}/\partial\varepsilon_b\approx +278$ K and
$\partial T_{BEC}/\partial\varepsilon_c\approx -1210$ K. The same
uniaxial strain derivatives of $T_{BEC}$ for the model lattice given
in Fig.1B at $\kappa^2/(2M\hbar\omega^3)=9.265$ are: $\partial
T_{BEC}/\partial\varepsilon_a\approx -115$ K and $\partial
T_{BEC}/\partial\varepsilon_c\approx -874$ K.

The obtained results clearly demonstrate a strong dependence of
$T_{BEC}$ on the arrangement of ions in the lattice. This has a
crucial effect on the value of $T_{BEC}$. The sign of $\partial
T_{BEC}/\partial\varepsilon$ is different for $a$- and $b$- axes,
which is caused by mutual arrangements of ions. In particular, for
the lattices in Fig.1A and Fig.1B one finds $\partial
T_{BEC}/\partial\varepsilon_b\approx +278$ K and $\partial
T_{BEC}/\partial\varepsilon_a\approx -115$ K, respectively. The
uniaxial strain derivatives, $\partial
T_{BEC}/\partial\varepsilon_c$, of the two lattices are both
negative. These results clearly show that  the two model lattices in
some ways qualitatively characterize the situations in
RBa$_2$Cu$_3$O$_{7-\delta}$ compounds under pressure (strain). Thus
compressive pressure (or strain) along the $b$-axis ($a$-axis)
increases (lowers) the value of $T_{BEC}$ in analogy with increase
(decrease) of $T_c$ of RBa$_2$Cu$_3$O$_{7-\delta}$ compounds under
compressive pressure ( or strain) in the same direction. The effect
of compressive pressure (strain) along the $c$-axis is similar to
that of along the $a$-axis. In terms of quantity, one should be
aware that our findings obtained relative to the lattices in Fig.1,
and not to real cuprates. Considering more real model structures
similar to the real structure of YBCO cuprates, one may obtain
better value of $\partial T_{BEC}/\partial\varepsilon_i$, close to
the $\partial T_c/\partial\varepsilon_i$ of YBCO. On the other hand,
the values of the uniaxial pressure (strain) derivative of $T_c$
along crystallographic axes $i=a,b,c$, measured, in different
experiments, are spread over a wide range, and in some cases
contradicts to each other. Welp et al. were the first to present
direct measurements of $\partial T_c/\partial p_i$ for the untwinned
YBa$_2$Cu$_3$O$_{7-\delta}$ single crystal \cite{welp-prl-69}. Their
results are: $\partial T_c/\partial p_a=-2.0\pm 0.2$ K/GPa,
$\partial T_c/\partial p_b=+1.9\pm 0.2$ K/GPa and $\partial
T_c/\partial p_c=-0.3\pm 0.1$ K/GPa. Bud'ko et al. obtained uniaxial
pressure (strain) derivatives of the critical temperature of
RBa$_2$Cu$_3$O$_{7-\delta}$ cuprate from the hydrostatic pressure
dependence, measured on the films of different crystalline
orientations \cite{budko}. According to Ref. \cite{budko}, $\partial
T_c/\partial p_a=-3.06\pm 0.35$ K/GPa ($\partial
T_c/\partial\varepsilon_a\approx -362\pm 50$ K) , $\partial
T_c/\partial p_b=+0.38\pm 0.18$ K/GPa ($\partial
T_c/\partial\varepsilon_b\approx +301\pm 30$ K) and $\partial
T_c/\partial p_c=+3.45\pm 0.43$ K/GPa ($\partial
T_c/\partial\varepsilon_c=+239\pm 24$ K). Pickett, in his paper
\cite{pickett}, quoting to the experimental results of Refs.
\cite{meingast-prl-91,welp-prl-69}, gives unexpected data: $\partial
T_c/\partial\varepsilon_a\approx +212$ K, $\partial
T_c/\partial\varepsilon_b\approx -244$ K and $\partial
T_c/\partial\varepsilon_c=-8$ K. One can see that our results are
close to the $\partial T_c/\partial\varepsilon_i$ of Ref.
\cite{budko}.

Now let's imagine that one has a hypothetic a quasi-two-dimensional
anisotropic lattice in which the interaction of charge carriers
(holes or electrons) with out-of-plane ions is strong, and that when
external pressure is applied along the $a$- ($b$-) axis this
interaction occurs in the analogous way to that as in the
one-dimensional lattice of Fig.1B (Fig.1A). Then, the
quasi-two-dimensional anisotropic lattice has all features of YBCO
cuprates with respect to the influence of uniaxial strain (pressure)
on $T_c$. Indeed, solving the system of equations
\begin{equation}\label{strain-pressure-system}
  \sum_jC_{ij}\frac{\partial T_{BEC}}{\partial p_j}=\frac{\partial
T_{BEC}}{\partial\varepsilon_i},
\end{equation}
with the set of elastic constants of the YBCO cuprate (all in GPa)
$C_{aa}=231$, $C_{ab}=132$, $C_{ac}=71$, $C_{bb}=268$, $C_{bc}=95$
and $C_{cc}=186$ taken from Ref. \cite{lei}, one finds
$\partial T_{BEC}/\partial
  p_a=-0.65$, $\partial T_{BEC}/\partial
  p_b=+3.58$ and $\partial T_{BEC}/\partial
  p_c=-6.27$ (all in K/GPa). Use of other set of elastic parameters (all in GPa) $C_{aa}=283$,
$C_{ab}=148$, $C_{ac}=83.1$, $C_{bb}=304$, $C_{bc}=109$
and $C_{cc}=236$ taken from Ref. \cite{ludwing-f-w}, yields  $\partial T_{BEC}/\partial
  p_a=-0.54$, $\partial T_{BEC}/\partial
  p_b=+2.91$ and $\partial T_{BEC}/\partial
  p_c=-4.86$ (all in
K/GPa). These findings indicate that the Bose-Einstein condensation
scenario of the ideal Bose-gas of intersite bipolarons is, {\it in
principle}, able to qualitatively explain the uniaxial strain
(pressure) experiments, regarding the effect of the strain
   (pressure) on $T_c$ of YBCO cuprates.
Quantitative discrepancies of our results and experimental data
may be the result of several factors: (i) the simplicity of the
model lattices under consideration. In reality one should consider more
complex structures than in Fig.1; (ii) the choice of the analytical
formula for the density-displacement type electron-lattice force;
(iii) the assumption that intersite bipolarons form an ideal
Bose-gas: in reality, due to other factors, there may be deviation
from the ideal case, leading to the formation of a
nonideal Bose-gas or Bose-liquid; (iv)
superconductivity of the YBCO cuprate may be due not only to
electron-phonon interaction, but may also have contributions from other
interactions as well. These factors suggest that we should  undertake more
comprehensive research on the studied problem.

The proposed model serves as a universal tool for studying strain
(pressure) induced effects in the cuprates. Its common features are
relevant to all cuprates.  In contrast with some theoretical
approaches (see for example \cite{klein-simanovsky}) the model
allows one to interpret the influence of pressure (strain) on
$T_{BEC}$ ($T_c$) along each axe, independently of the others.
Meanwhile, the model is able to account for interference of strains
between axes via the Poisson relation
$\nu=\varepsilon_a/\varepsilon_c$ or
$\nu=\varepsilon_b/\varepsilon_c$. This may be also useful in
theoretical studies of cuprate films grown on different substrates.

\section{Conclusion}

In conclusion, we have studied the effect of uniaxial strain
(pressure) on the temperature of Bose-Einstein condensation of
intersite bipolarons within the framework of the Extended Holstein
model. Uniaxial strain derivatives of $T_{BEC}$ are determined for
different lattices. It is found that $\partial
T_{BEC}/\partial\varepsilon_i$ depends strongly on the arrangement
of ions in the lattice. In particular, it may be positive or
negative. The results for the lattices under study (Fig.1) in some
way mimic the influence of uniaxial strain (pressure) on the
critical temperature, $T_c$, of YBCO cuprates. The calculated values
of the pressure derivatives of $T_{BEC}$ for the hypothetic lattice
qualitatively agree with the observed values of $\partial
T_c/\partial p_i$ ($i=a,b,c$) for YBCO cuprates.\\

{\bf Acknowledgement}\\

The authors wish to acknowledge Dr. P.R.Fraser and Dr. M.Ermamatov
for reading the manuscript and  helpful suggestions. The work was
done within the Fundamental Research Programme of Uzbek Academy of
Sciences (grant no. ${\Phi}$2-${\Phi}$A-${\Phi}$120) and was
partially supported by the Fund for Basic Research of Uzbek Academy
of Sciences (grant no. $\Phi$.2-12).


\begin{thebibliography}{99}
\bibitem{alekor}  {\small A.S.Alexandrov and P.E.Kornilovich, Phys. Rev. Lett. {\bf
82} (1999) 807.}
\bibitem{hol}  {\small T.Holstein, Ann. Phys. {\bf 8} (1959) 325; {\it %
ibid} {\bf 8} (1959) 343.}
\bibitem{devr-asa} {\small Jozef T. Devreese and Alexandre S. Alexandrov, Rep. Prog.
Phys. {\bf 72} (2009) 066501.}
\bibitem{kor-ctqmc} {\small P.E.Kornilovitch, Phys. Rev. Lett. {\bf 81} (1998) 5382.}
\bibitem{kor-giant} {\small P.E.Kornilovitch, Phys. Rev. B {\bf 59} (1999) 13531.}
\bibitem{kor-ground} {\small P.E.Kornilovitch, Phys. Rev. B {\bf 60} (1999) 3237.}
\bibitem{flw}  {\small H.Feshke, J.Loos, and G.Wellein, Phys. Rev. B {\bf\ 61}
(2000) 8016.}
\bibitem{pcf-jpcm}  {\small C.A.Perroni, V.Cataudella, and G. De Filippis, J. Phys.:
Condens. Matter {\bf 16} (2004) 1593.}
\bibitem{cfmp-prb}  {\small V.Cataudella, G. De Filippis, F.Martone and C.A.Perroni,
Phys. Rev. B {\bf 70} (2004) 193105.}
\bibitem{hohen} {\small Martin Hohenadler, Hans Gerd Evertz, and Wolfgang von der
Linden, Phys. Rev. B {\bf 69} (2004) 024301.}
\bibitem{asa-ya} {\small A.S.Alexandrov and B.Ya.Yavidov, Phys. Rev. B, {\bf 69}
(2004) 073101.}
\bibitem{trg}  {\small S.A.Trugman, J.Bon\v{c}a, and Li-Chung Ku, Int. J. Modern
Phys. B {\bf 15} (2001) 2707.}
\bibitem{yav-jetp} {\small B.Ya.Yavidov, Zh. Eksp. Teor. Fiz. {\bf 135} (6) (2009)
1173; JETP {\bf 108} (6) (2009) 1019.}
\bibitem{yav-physb} {\small Bakhrom Yavidov, Physica B, {\bf 404} (2009) 3756.}
\bibitem{spencer} {\small P.E.Spencer, J.H.Samson, P.E.Kornilovitch, and
A.S.Alexandrov, Phys. Rev. B, {\bf 71} (2005) 184310.}
\bibitem{hague-etal} {\small J.P.Hague, P.E.Kornilovitch, A.S.Alexandrov, and
J.H.Samson, Phys. Rev. B {\bf 73} (2006) 054303.}
\bibitem{hague-kor} {\small J.P.Hague and P.E.Kornilovitch, Phys. Rev. B {\bf 80}
(2009) 054301.}
\bibitem{yav-pla} {\small B.Ya.Yavidov, Sh.S. Djunamov and S.Dzhumanov, Physics
Letters A {\bf 374} (2010) 2772.}
\bibitem{yav-epjb} {\small B.Ya.Yavidov, Eur. Phys. J. B {\bf 75} (2010) 481.}
\bibitem{stojan} {\small V.M.Stojanovi\'{c}, P.A.Bobbert and M.A.J.Michels, Phys.
Stat. Sol. C {\bf 1} (2004) 172.}
\bibitem{meisel} {\small K.D.Meisel, H.Vocks, P.A.Bobbert, Phys. Rev. B {\bf 71}
(2005) 205206.}

\bibitem{yav-physc-strain} {\small B.Ya. Yavidov, Physica C {\bf 471} (2011) 71.}
\bibitem{sato-naito} {\small H.Sato and M.Naito, Physica C {\bf 274} (1997) 221.}
\bibitem{locquet} {\small J.-P.Locquet, J.Perret, J.Fompeyrine, E.M\"{a}chler,
J.W.Seo, and G.Van Tendeloo, Nature {\bf 394} (1998) 453}
\bibitem{schilling-handbook}    {\small J.S.Schilling, in Handbook of
High-temperature Superconductivity. Theory and Experiment,
edited by J.Robert Schrieffer (Springer-Verlag, New York, 2007),
p.427}

\bibitem{millis-rabe} {\small A.J. Millis, K.M. Rabe, Phys. Rev. B 38 (1988) 8908.}
\bibitem{q.p.li} {\small Q.P. Li, Physica C 209 (1993) 513.}
\bibitem{klein-simanovsky} {\small Michael W. Klein and Sergey B. Simanovsky, Phys.
Rev. Lett. 78 (1997) 3927.}
\bibitem{pickett} {\small W.E. Pickett,Physica C {\bf 289} (1997) 51.}
\bibitem{jansen-block} {\small L. Jansen and R. Block, Physica A {\bf 264} (1999) 523.}
\bibitem{chen-habermeier} {\small X.J. Chen, H.Q. Lin, W.G. Yin, C.D. Gong, and
H.-U. Habermeier, Phys. Rev. B {\bf 64} (2001) 212501.}

\bibitem{meingast-prl-91} {\small C. Meingast, O. Kraut, T. Wolf, H. W\"{u}hl, A.
Erb and G. M\"{u}ller-Vogt, Phys. Rev. Lett. {\bf 67} (1991) 1634.}
\bibitem{welp-prl-69} {\small U. Welp, M. Grimsditch, S. Fleshler, W. Nessler, J.
Downey, G.W. Crabtree and J. Guimpel, Phys. Rev. Lett. {\bf 69} (1992) 2130.}

\bibitem{gunnar-rosch-jpcm-2008} {\small O. Gunnarsson and O. R\"{o}sch, J.Phys.:
Condens. Matter {\bf 20} (2008) 043201.}
\bibitem{mish-phys.usp-2009} {\small A.S. Mishchenko, Phys. Usp. {\bf 52} (2009) 1193.}

\bibitem{zhao-Pol.Adv.Mat} {\small Guo-meng Zhao, in {\it Polarons in Advanced
Materials} Edited by A.S.Alexandrov (Springer, 2007) p.569}
\bibitem{BussHol-Keller-Pol.Adv.Mat} {\small Annette Bussmann-Holder and Hugo
Keller, in {\it Polarons in Advanced Materials} Edited by A.S.Alexandrov (Springer,
2007) p.599}

\bibitem{asa-kor-jpcm}  {\small A.S.Alexandrov and P.E.Kornilovich, J. Phys.:
Condens. Matter. {\bf 14} (2002) 5337.}

\bibitem{timusk}    {\small T.Timusk, C.C.Homes, and W.Reichardt, in Anharmonic
Properties of High-$T_c$ Cuprates
, edited by D.Mihailovi\'{c} et. al. (Wolrd Scientific, Singapore,
1995), p.171}
\bibitem{bt}  {\small J.Bonca and S.A.Trugman, Phys. Rev. B {\bf\ 64} (2001) 094507.}
\bibitem{lang-fir}  {\small I.G.Lang, Yu.A. Firsov, Zh. Eksp. Teor. Fiz. {\bf 43}
(1962) 1843, Sov. Phys.-JETP {\bf 16} (1963) 1301.}

\bibitem{budko} {\small S.L. Bud'ko, J. Guimpel, O. Nakamura, M.B. Maple and Ivan K.
Schuller, Phys. Rev. B 46 (1992) 1257.}

\bibitem{lei} {\small M. Lei et al., Bull. Am. Phys. Soc. 37 (1992) 649.}
\bibitem{ludwing-f-w} {\small H.A. Ludwing, W.H. Fietz and H.W\"{u}hl, Physica C 197
(1992) 113.}




\end{thebibliography}
\end{document}